\documentclass{article}%
\usepackage[top=3cm, bottom=3cm, left=3cm, right=3cm]{geometry}
\usepackage{amsmath}
\usepackage{amsfonts}
\usepackage{amssymb}
\usepackage{graphicx}%
\begin{document}

\title{Decay law and time dilatation}
\author{Francesco Giacosa\\\textit{Institute of Physics, Jan-Kochanowski University, }\\\textit{ul. Swietokrzyska 15, 25-406, Kielce, Poland.}\\\textit{Institute for Theoretical Physics, J. W. Goethe University, }\\\textit{ Max-von-Laue-Str. 1, 60438 Frankfurt, Germany.}}
\date{}
\maketitle

\begin{abstract}
We study the decay law for a moving unstable particle. The usual
time-dilatation formula states that the decay width for an unstable state
moving with a momentum $p$ and mass $M$ is $\tilde{\Gamma}_{p}=\Gamma
M/\sqrt{p^{2}+M^{2}}$ with $\Gamma$ being the decay width in the rest frame.
In agreement with previous studies, we show that in the context of QM as well
as QFT this equation is \textit{not} correct provided that the quantum
measurement is performed in a reference frame in which the unstable particle
has momentum $p$ (note, a momentum eigenstate is \textit{not} a velocity
eigenstate in QM). We then give, to our knowledge for the first time, an
analytic expression of an improved formula and we show that the deviation from
$\tilde{\Gamma}_{p}$ has a maximum for $p/M=\sqrt{2/3},$ but is typically
\textit{very} small. Then, the result can be easily generalized to a momentum
wave packet and also to an arbitrary initial state. Here, we give a very
general expression of the non-decay probability. As a next step, we show that
care is needed when one makes a boost of an unstable state with zero
momentum/velocity: namely, the resulting state has zero overlap with the
elements of the basis of unstable states (it is already decayed!). However,
when considering a spread in velocity, one finds again that $\tilde{\Gamma
}_{p}$ is typically a very good approximation. The study of a velocity
wave-packet represents an interesting subject which constitutes one of the
main outcomes of the present manuscript.\textbf{ }In the end, it should be
stressed that there is no whatsoever breaking of special relativity, but as
usual in QM, one should specify which kind of measurement on which kind of
state is performed.

\end{abstract}

\section{Introduction}

Time dilatation is one of the most striking and beautiful consequences of
special (and also general) relativity. It is experimentally precisely
verified, yet it still fascinates us because it is so much different from our
common sense: it shows the inexistence of a `present' for all the observers
and it generates many `paradoxes', such as the renowned twins' one. Its
formulation is now parts of textbooks. Two observes, Alice (A) and Bob (B),
move away from each other with constant velocity $v$. The relation among the
coordinates of Bob and Alice is given by a Lorentz transformation. Alice has a
clock (of whatever type!) in her hands. The clock makes a `click' after a
certain time $T$ which links the events $(0,0)$ and $(T,0)$ in her reference
frame. Then, the time interval between the two events as measured by Bob is
$\gamma T$ with $\gamma=1/\sqrt{1-v^{2}}.$ This is the usual time dilatation.

Another fascinating -- and in a certain sense even more astonishing-- part of
modern physics is Quantum Mechanics (QM), in which randomness plays a central
role: when a measurement is performed, the wave function of a quantum state
instantaneously collapses to one of the eigenstates of the measured
observable. An unstable quantum state, such as an unstable elementary
particle, is one of the best examples to show the features of QM: the state of
the system is a superposition of decayed and undecayed, and --upon measurement
through an appropriate apparatus-- a collapse to either decayed or not decayed
takes place. Indeed, Schr\"{o}dinger \cite{schroedinger} had chosen the decay
of a radioactive atom to trigger his sadistic but fortunately only ideal
cat-killing machine, in such a way to put also the state of the cat into a
superposition of alive and dead (unless, of course, the cat itself is regarded
as a measuring device). The merging of special relativity and QM is
Quantum\ Field Theory (QFT), which is the correct theoretical framework to
describe to creation and annihilation of particles and thus -in ultimate
analysis- to describe decays.

Now, Alice can use unstable particles as a clock. She prepares at $t=0$ a box
of $N_{0}$ unstable states which have zero velocity, i.e. they are all at rest
with Alice. She performs measurements and finds that the number of states
decreases as $N(t)=N_{0}P_{nd}(t),$ where $P_{nd}(t)$ is the nondecay
probability as measured by Alice. In the exponential limit \cite{ww,scully}:
\begin{equation}
P_{nd}(t)=e^{-\Gamma t}\text{ , }\tau=\Gamma^{-1}\text{ `lifetime'.}
\label{alice}%
\end{equation}
It is now theoretically
\cite{khalfin,misra,dega,ghirardi,koshinorev,facchiprl,zenoqft,duecan} and
experimentally \cite{raizen,rothe} established that deviations from the
exponential decay exist, but they are usually small. We shall consider at
first this limit in our numerical examples, but we will also present general
expressions which naturally take into account departures from the exponential decay.

Now, what about Bob? The usual answer is that Bob `sees' the unstable
particles living longer. This is because of time dilatation. Thus, Bob should
`see' an increased lifetime of $\tau_{B}=\gamma\tau>\tau.$ In other words, the
decay width as measured by Bob is given by the standard \textquotedblleft
Einstein\textquotedblright\ formula:%
\begin{equation}
\frac{\Gamma}{\gamma}\equiv\frac{\Gamma M}{\sqrt{p^{2}+M^{2}}}=\tilde{\Gamma
}_{p}\text{ ,} \label{gammatilde}%
\end{equation}
where $M$ is the rest mass of the unstable particle. This is indeed the usual
explanation behind the famous expression `the long lifetime of the muons'.
Namely, the muons produced in the atmosphere manage to reach detectors located
on the earth because of time dilatation: they move fast enough so that, in the
earth reference frame, the decay width is sizably smaller and the decay time
larger (see also the direct experimental verification in Ref. \cite{bailey}).
But what does it mean to `see'? We need to be precise in this respect.

Namely, it is important to stress from the very beginning the following point:
if Alice performs the decayed/undecayed measurement in her reference frame,
then Bob confirms the formula (\ref{gammatilde}) \emph{exactly}. The question
here is slightly but crucially different: what does it happen if the detector
is in the rest frame of Bob, who performs the decayed/undecayed measurement?
In other words, is there a difference if it is Alice or Bob who is making the experiment?

We study the problem in four different steps. First (Sec. 2) we consider the
case in which the unstable state has a definite momentum $p$ w.r.t. the
measuring device. Here, we review previous existing works on the subject
\cite{exner,khalfin2,shirokovold,shirokov,shirokovnew,stefanovich,stefanovichnew,stefanovichbook,urbanowski}%
: the general outcome is that Eq. (\ref{gammatilde}) does not hold for such a
system, even if it is a very good approximation. Moreover, we also present an
analytical formula for the decay width $\Gamma_{p}$ in the Breit-Wigner limit.
At the same time, we quantify the deviations in Fig. 1 and give further
analytic expressions for the deviations from Eq. (\ref{gammatilde}). In this
context, it is important to stress that in QM and QFT an unstable state with
definite momentum does not have a definite velocity, thus this situation does
not correspond to a boost but rather to a momentum translation of a
zero-velocity unstable state. Then,\ strictly speaking, this is not what Bob
would measure, but it represents a very well defined theoretical setup (for
instance, for a \textquotedblleft momentum translated\textquotedblright%
\ observer Charlie).

The second step (Sec. 3) is the generalization to a momentum wave packet. Here
a subtle but very important point concerns the definition of the basis of
unstable states: we shall argue that the nondecay probability of a generic
state $\left\vert \Psi\right\rangle $ is not the survival probability
$\left\vert \left\langle \Psi\left\vert e^{-iHt}\right\vert \Psi\right\rangle
\right\vert ^{2}$, but that one should project onto a suitable basis. We also
argue that the only consistent choice of such a basis is to use the set of
unstable states with definite momentum (which can be obtained from the
zero-velocity unstable state by momentum translations). Then, in the end of
this section we also study the non-decay probability for a fully arbitrary
initial state, leading to the most general result of the present paper.

As a third step (Sec. 4) we study -by using the formalism developed in Sec. 3-
a boost of a zero-velocity unstable state and thus obtain an eigenstate of
velocity (this is what Bob would measure). Quite surprisingly, the boosted
state is already decayed: the boosted muon consists -for Bob and for his
measuring apparatus- of an electron and two neutrinos; a boosted pion consists
of two photons, and so on and so fort. This result is in agreement with Ref.
\cite{stefanovichnew}, where it was called `decay caused by boost'. In this
respect, Eq. (\ref{gammatilde}) is completely wrong. In this section we also
discuss the survival probability of a state with definite velocity, which -as
presented in Refs. \cite{stefanovichnew,giunti,shirokovnew}- shows a quite
amusing Lorentz contraction of time instead of dilation. This problem does not
arise if the definition of Sec. 3 concerning the decayed/undecayed assessment
is used.

The result of Sec. 4 is however very `delicate' as we describe in the fourth
and last step (Sec. 5): it is strictly valid only if we perform a boost on a
state which have exactly zero momentum (or velocity). If a more realistic
velocity wave-packet is considered, we find again that the decay law is well
approximated by Eq. (\ref{gammatilde}). Namely, using again Sec. 3, there is a
nonzero overlap with the set of undecayed states as soon as a velocity
spreading is implemented. The results, which are depicted in\ Fig. 2,
constitute one of the main contributions of the present paper. Finally, we
summarize the manuscript in\ Sec. 6.

\section{Decay law of a state with definite momentum}

We consider a spacial one-dimensional system described by\emph{ }the
Hamiltonian $H$, whose eigenstates are denoted as%
\[
\left\vert m,p\right\rangle =U_{p}\left\vert m,0\right\rangle \text{ ,}%
\]
where $U_{p}$ is the unitary operator associated to the translation in
momentum space. The state $\left\vert m,p\right\rangle $ has definite energy,
$H\left\vert m,p\right\rangle =\sqrt{p^{2}+m^{2}}\left\vert m,p\right\rangle
$, definite momentum, $P\left\vert m,p\right\rangle =p\left\vert
m,p\right\rangle ,$ as well as a definite velocity: $\left(  P/H\right)
\left\vert m,p\right\rangle =\left(  p/\gamma m\right)  \left\vert
m,p\right\rangle .$ The normalization is given by: $\left\langle m_{1}%
,p_{1}|m_{2},p_{2}\right\rangle =\delta(m_{1}-m_{2})\delta(p_{1}-p_{2}).$

An unstable state with definite momentum $p$ is denoted as $\left\vert
S,p\right\rangle =U_{p}$ $\left\vert S,0\right\rangle $:%
\begin{equation}
\left\vert S,p\right\rangle =\int_{0}^{\infty}\mathrm{dm}a_{S}(m)\left\vert
m,p\right\rangle \text{ .} \label{sp}%
\end{equation}
The quantity $d_{S}(m)=\left\vert a_{S}(m)\right\vert ^{2}$ is the mass
distribution: $d_{S}(m)dm$ is the probability that the unstable particle $S$
has a mass between $m$ and $m+dm.$ As a consequence, $\int_{0}^{\infty}%
dmd_{S}(m)=1$. The normalization $\left\langle S,p_{1}|S,p_{2}\right\rangle
=\delta(p_{1}-p_{2})$ follows. Note, Eq. (\ref{sp}) is not an state with
definite velocity. This is due to the fact that each state $\left\vert
m,p\right\rangle $ in the superposition has a different velocity $p/\left(
\gamma m\right)  $. The theory of decays is discussed in great detail for the
case $p=0$ in Refs. \cite{ghirardi,lee,ford} and Refs. therein. Moreover, as
shown in\ Ref. \cite{duecan}, QFT at the one-loop resummed level can be also
described within such formalism. Thus, our discussions includes from the very
beginning not only QM but also QFT.

A very useful, although not exact, approximation both in QM and in QFT is the
Breit-Wigner (BW) formula, $d_{S}^{BW}(m)=\frac{\Gamma}{2\pi}\left[
(m-M)^{2}+\Gamma^{2}/4\right]  ^{-1}$. By starting from a properly normalized
state with zero momentum, $\left\vert S,0\right\rangle /\sqrt{\delta(p=0)}$
and upon extending the integral to negative masses (this is strictly speaking
unphysical, but the induced error is very small if the ratio $\Gamma/M\ll1$)
we obtain the amplitude
\begin{equation}
a(t,0)=\frac{1}{\delta(p=0)}\left\langle S,0\left\vert e^{-iHt}\right\vert
S,0\right\rangle \simeq\int_{-\infty}^{\infty}\mathrm{dm}d_{S}^{BW}%
(m)e^{-imt}=e^{-iMt-\Gamma t/2}\text{ ,} \label{at0}%
\end{equation}
ergo the survival probability of the state $\left\vert S\right\rangle $ is
$P_{nd}(t)=\left\vert a(t,0)\right\vert ^{2}=e^{-\Gamma t}$, which is the
usual decay law already introduced in Eq. (\ref{alice}).

As mentioned in the Introduction, it is now well known that deviations from
the exponential decay exist. They stem from the fact that any realistic
distribution $d_{S}(m)$ should be zero below a certain energy threshold
$m_{0}\geq0$, thus the integral in the r.h.s. of Eq. (\ref{at0}) should extend
from $m_{0}$ upward (e.g., in a two-body decay the threshold energy is given
by the sum of the rest masses produced in the decay, see also\ Ref.
\cite{shirokov} for the study of the useful case $m_{0}=0$). Moreover,
$d_{S}(m)$ should also fall down faster than $1/m^{2}$ for large $m$ due to
form factors \cite{ghirardi}. Both properties lead to long-time and short-time
deviations from the exponential decay law $P_{nd}(t)=\left\vert
a(t,0)\right\vert ^{2}=e^{-\Gamma t},$ respectively. Yet, these deviations are
usually small and occur in a very short time interval at the beginning of the
decay and at a very late decay times (typically, of the order of
$10\Gamma^{-1}$). Thus, there is a long time interval in which the exponential
decay is very well realized. In this work, we aim to concentrate on the study
of decays in relation with special relativity. Thus, in the present paper, we
shall limit our numerical investigations to the Breit-Wigner limit. The reason
is that, as we shall see, there are -even in this limit- interesting aspects
which need further investigations and clarifications. Yet, general expressions
valid beyond the Breit-Wigner limit are presented in the next section.

We now turn to the calculation of the non-decay probability of a state with
definite but non vanishing momentum and present a derivation similar to the
works of Refs.
\cite{khalfin2,shirokovold,shirokov,shirokovnew,stefanovich,stefanovichnew,stefanovichbook,urbanowski}%
. To this end, we consider a (normalized) state $\left\vert S,p\right\rangle
/\sqrt{\delta(p=0)}$ with a generic momentum $p,$ we find the following
non-decay amplitude:
\begin{align}
a(t,p)  &  =\frac{1}{\delta(p=0)}\left\langle S,p\left\vert e^{-iHt}%
\right\vert S,p\right\rangle =\int_{-\infty}^{\infty}\mathrm{dm}%
d_{S}(m)e^{-i\sqrt{m^{2}+p^{2}}t}\nonumber\\
&  \simeq\int_{-\infty}^{\infty}\mathrm{dm}d_{S}^{BW}(m)e^{-i\sqrt{m^{2}%
+p^{2}}t}=e^{-i\sqrt{(M-i\Gamma/2)^{2}+p^{2}}t}\text{ .} \label{atp}%
\end{align}
For the investigation of similar equations we refer also to Refs.
\cite{shirokov,stefanovich,urbanowski}. In the BW-limit, the non-decay
(survival) probability reads%
\begin{equation}
P_{nd}(t)=\left\vert a(t,p)\right\vert ^{2}=e^{-\Gamma_{p}t}%
\end{equation}
with
\begin{equation}
\Gamma_{p}=2\operatorname{Im}\left[  \sqrt{(M-i\Gamma/2)^{2}+p^{2}}\right]
\text{ .}%
\end{equation}
Quite interesting, $\Gamma_{p}$ can be recasted in the analytic expression:
\begin{equation}
\Gamma_{p}=\sqrt{2}\sqrt{\left[  \left(  M^{2}-\frac{\Gamma^{2}}{4}%
+p^{2}\right)  ^{2}+M^{2}\Gamma^{2}\right]  ^{1/2}-\left(  M^{2}-\frac
{\Gamma^{2}}{4}+p^{2}\right)  }\text{ .} \label{gammap}%
\end{equation}
One realizes that $\Gamma_{p}$ \textit{differs} from the standard formula of
Eq. (\ref{gammatilde}): $\Gamma_{p}\neq\tilde{\Gamma}_{p}=\Gamma M/\sqrt
{p^{2}+M^{2}}$. Deviations where already discussed in Refs.
\cite{exner,khalfin2,shirokovold,shirokov,shirokovnew,stefanovich,stefanovichnew,stefanovichbook,urbanowski}%
. We thus confirm those results and present -to our knowledge for the first
time- the analytic expression (\ref{gammap}) for the decay width of an
unstable particle with definite momentum $p$. One can indeed easily proof that
for $\Gamma/M\ll1$ one reobtains, as expected, that $\Gamma_{p}\simeq
\tilde{\Gamma}_{p}$. Notice that, while our Eq. (\ref{gammap}) is only valid
in the (typically very good) exponential limit, we feel that it is always
useful to have -whenever possible- an analytic expression, especially in the
context of connecting Quantum Mechanics and Special\ Relativity.

As a next point, it is natural to ask how good Eq. (\ref{gammatilde}) is. For
small $\Gamma/M$, the difference
\begin{equation}
\Delta=\frac{\Gamma_{p}-\tilde{\Gamma}_{p}}{M}%
\end{equation}
has a maximum for
\begin{equation}
\frac{p_{\max}}{M}=\sqrt{\frac{2}{3}}\simeq0.816\text{ .} \label{pmax}%
\end{equation}
The value of the difference at the maximum scales as:%
\begin{equation}
\Delta_{\max}=\frac{\Gamma_{p_{\max}}-\tilde{\Gamma}_{p_{\max}}}{M}\simeq
\frac{3}{100}\sqrt{\frac{3}{5}}\left(  \frac{\Gamma}{M}\right)  ^{3}\text{ .}
\label{deltamax}%
\end{equation}
As an example, in Fig. 1 we plot $\Delta$ as function of $p/M$ for the
numerical choice $\Gamma/M=1/100$.

Eq. (\ref{gammatilde}) turns out to be an \textit{extremely} good
approximation. For instance, in the case of the muon ($M=105.6583$ MeV,
$\Gamma=2.99\cdot10^{-16}$ MeV), we get an astonishingly small deviation:
$\Gamma_{p_{\max}}-\tilde{\Gamma}_{p_{\max}}\simeq5.598\cdot10^{-53}$ MeV. The
deviation increases only very slightly with increasing decay width. For the
neutral pion ($M=134.9766$ MeV, $\Gamma=7.72\cdot10^{-6}$ MeV) we get
$\Gamma_{p_{\max}}-\tilde{\Gamma}_{p_{\max}}\simeq5.81\cdot10^{-22}$ MeV. Only
for strong decays the deviations become somewhat larger. For the $\rho$ meson,
($M=775.26$ MeV, $\Gamma=147.8$ MeV) we get $\Gamma_{p_{\max}}-\tilde{\Gamma
}_{p_{\max}}\simeq0.125$ MeV (all numerical values are from Ref. \cite{pdg}).

\bigskip%

\begin{figure}
[ptb]
\begin{center}
\includegraphics[
height=3.3114in,
width=5.6083in
]%
{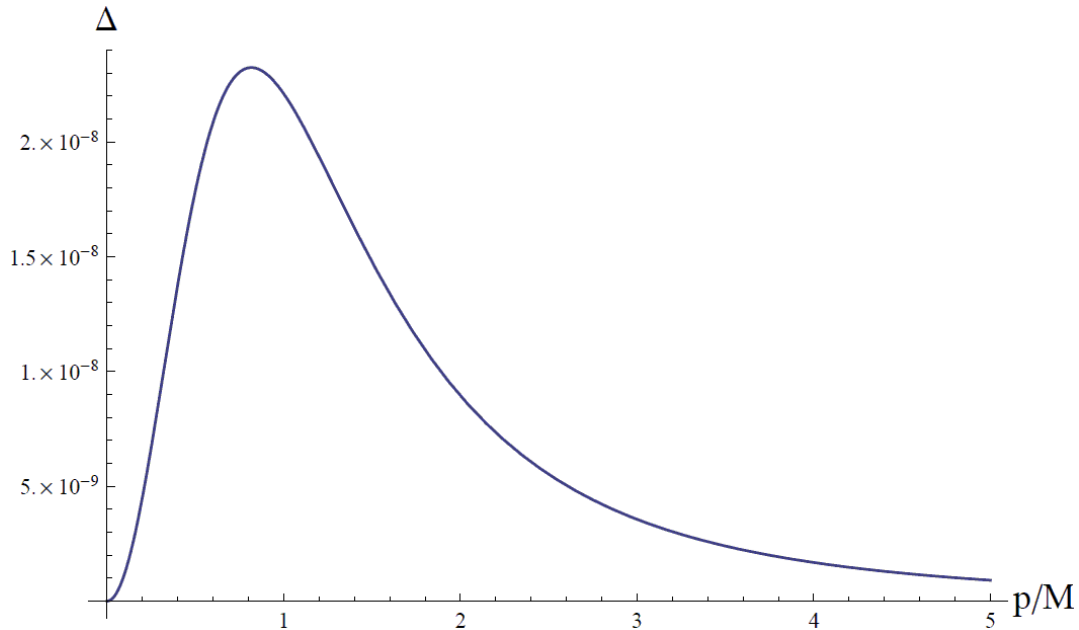}%
\caption{Difference between our result of Eq. (\ref{gammap}) and the standard
formula of Eq. (\ref{gammatilde}) as function of $p$ for the ratio
$\Gamma/M=1/100.$ The position of the minimum $p_{\max}/M=\sqrt{2/3}$ is
constant for decreasing $\Gamma/M.$ The height of the maximum scales as given
in Eq. (\ref{deltamax}), thus the deviation decreases very fast for decreasing
$\Gamma/M.$}%
\end{center}
\end{figure}

A direct experimental verification of this deviation seems at present
difficult because of the very small departure (a very high precision is
required, much beyond the one in Ref. \cite{bailey}). However, it is a very
interesting deviation from a fundamental point of view.

One may wonder if there is a conflict with the usual textbook QFT results in
which the decay width in QFT is expressed in a covariant form, see for example
Ref. \cite{pdg,peskin} for the two-body decay:
\begin{equation}
d\Gamma=\frac{(2\pi)^{4}}{2M}\left\vert \mathcal{M}\right\vert ^{2}%
\delta(p-k_{1}-k_{2})\frac{d^{3}k_{1}}{(2\pi)^{3}2E_{1}}\frac{d^{3}k_{2}%
}{(2\pi)^{3}2E_{2}} \label{qfts}%
\end{equation}
where $k_{i}=(E_{i},\vec{k}_{i})$ is the fourmomentum of the outgoing particle
$i$ with $i=1,2.$ The amplitude $\mathcal{M}$ is a Lorentz scalar. Then, the
kinematic factors reproduce Eq. (\ref{gammatilde}) exactly.\ The reason is
that this derivation of QFT is based on the S-matrix formalism, which in
designed to deal with scattering of asymptotic stable states. It can be
applied to decays, but with due care \cite{peskin}. Namely, Eq. (\ref{qfts})
is only valid in the limit $\Gamma$ $\ll M$ in which the unstable state can be
approximately considered as stable (in this limit, the mass and the momentum
of the unstable state are uniquely defined). Thus, the usual textbook QFT
expression is also a (typically very useful) approximation. Indeed, for the
very same reason, also deviations form the exponential decay law
\textit{cannot} be obtained in the S-matrix formalism of QFT but need more
advance approaches, see Ref. \cite{zenoqft,duecan}.

\section{Spreading of momentum (wave packet) and generalization}

An initial state with definite momentum is a (useful) idealization. A more
realistic initial state is expressed by a superposition of different states of
the type $\left\vert S,p\right\rangle $:%
\begin{equation}
\left\vert \Psi\right\rangle =\int_{-\infty}^{+\infty}\mathrm{dp}%
B(p)\left\vert S,p\right\rangle \text{ with }\int_{-\infty}^{+\infty
}\mathrm{dp}\left\vert B(p)\right\vert ^{2}=1\text{ .} \label{pwp}%
\end{equation}
Which is the non-decay probability associated to this state? This point is
delicate: the quantity $\left\langle \Psi\left\vert e^{-iHt}\right\vert
\Psi\right\rangle $ is \textit{not} what we are looking for. Namely, this is
the probability's amplitude that the state is still in the very same initial
form, but it does not take into account the possibility that the state has
evolved into a different non-decayed state (a dephasing, but not a decay, may
occur). In order to obtain the non-decay probability we have to project onto
the basis of undecayed states $\left\vert S,p\right\rangle $. Thus, we first
calculate the amplitude that at the time $t$ the state is given by $\left\vert
S,p\right\rangle ,$ expressed by $\left\langle S,p\left\vert e^{-iHt}%
\right\vert \Psi\right\rangle =B(p)a(t,p)$, and then we square the modulus and
integrate over $p$. In the end, the non-decay probability reads:%
\begin{equation}
P_{nd}(t)=\int_{-\infty}^{+\infty}\mathrm{dp}\left\vert \left\langle
S,p\left\vert e^{-iHt}\right\vert \Psi\right\rangle \right\vert ^{2}\text{ .}
\label{pndgen}%
\end{equation}
\textbf{ }In other words, the space of undecayed states is not one-dimensional
but is spanned by all states $\left\vert S,p\right\rangle $, where $p$ is a
continuous variable. Only when the wave-packet is sufficiently narrow, we
recover Eq. (\ref{atp}), see also the discussion in Ref. \cite{exner}.
Moreover, only for a very narrow wave-packet the quantity $\left\vert
\left\langle \Psi\left\vert e^{-iHt}\right\vert \Psi\right\rangle \right\vert
^{2}$ coincides with the nondecay probability of Eq. (\ref{pndgen}).
Explicitly, the non-decay probability associated to the wave-packet of Eq.
(\ref{pwp}) reads (in the Breit-Wigner limit):%
\begin{equation}
P_{nd}(t)=\int_{-\infty}^{+\infty}\mathrm{dp}\left\vert B(p)\right\vert
^{2}\left\vert a(t,p)\right\vert ^{2}=\int_{-\infty}^{+\infty}\mathrm{dp}%
\left\vert B(p)\right\vert ^{2}e^{-\Gamma_{p}t}\simeq\int_{-\infty}^{+\infty
}\mathrm{dp}\left\vert B(p)\right\vert ^{2}e^{-\tilde{\Gamma}_{p}t}\text{ .}
\label{pndpwp}%
\end{equation}
We thus have a decay law which emerges as the sum of more exponential
functions, each of them weighted with the corresponding probability
$\left\vert B(p)\right\vert ^{2}$. The condition $P_{nd}(0)=1$ holds: the
wave-packet of Eq. (\ref{pwp}) is a non-decayed state at $t=0.$\textbf{ } For
$\left\vert B(p)\right\vert ^{2}=\delta(p-p_{0})/\delta(0)$ we reobtain the
decay law for a definite momentum $p_{0},$ see Eq. (\ref{gammap}). The last
expression in the right is what one could intuitively write down by using Eq.
(\ref{gammatilde}). In general, if $B(p)$ is centered around a certain value
$p_{0},$ the non-decay probability can be very well approximated by
$P_{nd}(t)\simeq e^{-\Gamma_{p_{0}}t}\simeq e^{-\tilde{\Gamma}_{p_{0}}t}$.

For our discussion it was not necessary to involve the space variable $x$.
Anyway, for completeness, the space-like wave function of the unstable state
$\left\vert \Psi\right\rangle $ can be easily obtained as:
\begin{equation}
\psi(t,x)=\left\langle x\left\vert e^{-iHt}\right\vert \Psi\right\rangle
=\frac{1}{2\pi}\int_{-\infty}^{+\infty}\mathrm{dp}B(p)a(t,p)e^{ipx}%
\end{equation}
\newline One has a wave-packet that is moving in space and that is gradually
loosing its overall normalization because of the decay.

As a last step, we consider the most general form of an initial state
$\left\vert \Theta\right\rangle $ as given by:
\begin{equation}
\mathbf{\ }\left\vert \Theta\right\rangle =\int_{0}^{\infty}\mathrm{dm}%
\int_{-\infty}^{+\infty}\mathrm{dp}\beta(m,p)\left\vert m,p\right\rangle
\text{ with }\int_{0}^{\infty}\mathrm{dm}\int_{-\infty}^{+\infty}%
\mathrm{dp}\left\vert \beta(m,p)\right\vert ^{2}=1\text{ .}\label{theta}%
\end{equation}
Namely, being the states $\left\vert m,p\right\rangle $ a basis for the whole
space (decayed and non-decayed), $\ \left\vert \Theta\right\rangle $ is the
most general form of a ket. The wave-packet $\left\vert \Psi\right\rangle $ of
Eq. (\ref{pndpwp}) is a special case obtained by setting $\beta(m,p)=B(p)a_{S}%
(m),$ i.e. a factorization of momentum and mass parts is present (as in\ Ref.
\cite{exner}). The (normalized) state $\left\vert S,p_{0}\right\rangle
/\sqrt{\delta(0)}$ is obtained for $\beta(m,p)=a_{S}(m)\delta(p-p_{0}%
)/\sqrt{\delta(0)}.$ The non-decay probability associated to the most general
initial state$\ \left\vert \Theta\right\rangle $ reads:%
\begin{equation}
\mathbf{\ }P_{nd}(t)=\int_{-\infty}^{+\infty}\mathrm{dp}\left\vert
\left\langle S,p\left\vert e^{-iHt}\right\vert \Theta\right\rangle \right\vert
^{2}\mathbf{\ }=\int_{-\infty}^{+\infty}\mathrm{dp}\left\vert \int_{0}%
^{\infty}\mathrm{dm}a_{S}^{\ast}(m)\beta(m,p)e^{-i\sqrt{m^{2}+p^{2}}%
t}\right\vert ^{2}.\label{pndtheta}%
\end{equation}
Here, $P_{nd}(0)\leq1$: the initial state is not necessarily a non-decayed
state. Notice that the present form of $P_{nd}(t)$ represents a generalization
of the discussion of Ref. \cite{stefanovichnew} and Ref. \cite{shirokovnew}.
We will use this Eq. (\ref{pndtheta}) also in Sec. 4 and 5 when studying boosts.

\section{State with a definite velocity}

The state $\left\vert S,p\right\rangle $ discussed in Sec. 1 has not a
definite speed. We can however construct a state with definite speed upon
boosting the state $\left\vert S,0\right\rangle $ :
\begin{equation}
U_{v}\left\vert S,0\right\rangle \equiv\left\vert S,v\right\rangle =\int
_{0}^{\infty}\mathrm{dm}a_{S}(m)\sqrt{m}\gamma^{3/2}\left\vert m,m\gamma
v\right\rangle \text{ .}\label{sv}%
\end{equation}
In this way, $\left\vert S,v\right\rangle $ is an eigenstate of the velocity
operator $P/H$ with eigenvalue $v$. The additional factors are needed to
obtain the normalization $\left\langle S,v_{1}|S,v_{2}\right\rangle
=\delta(v_{1}-v_{2}).$ It is indeed a straightforward but subtle link between
different bases. Note, one immediately sees from Eq. (\ref{sv}) that $m$ needs
to be positive. A normalized state with definite velocity is then given by
$\delta(v=0)^{-1/2}\left\vert S,v\right\rangle $; this state can be also
obtained from the general expression (\ref{theta}) upon setting\textbf{
}$\beta(m,p)=\delta(v=0)^{-1/2}\sqrt{m}\gamma^{3/2}a_{S}(m)\delta(p-m\gamma
v).$ This is indeed the state that Bob `sees' if a normalized state with
definite momentum/velocity $p=v=0$ is prepared by Alice. In order to evaluate
the non-decay probability associated to this state we first calculate the
amplitude
\begin{equation}
\left[  \delta(v=0)\right]  ^{-1/2}\left\langle S,v\left\vert e^{-iHt}%
\right\vert S,p\right\rangle =\left[  \delta(v=0)\right]  ^{-1/2}d_{S}\left(
\frac{p}{\gamma v}\right)  \sqrt{\frac{p}{v}}\frac{1}{v}e^{-i\gamma
mt}\label{noov}%
\end{equation}
Then, upon squaring and integrating over $p$, we find that $P_{nd}(t)=0$ for
any time! Namely, even at $t=0$ one has $P_{nd}(0)=0.$ This result can be also
obtained by using Eq. (\ref{pndtheta}). In other words, a boost transforms an
undecayed particle into decayed ones. In the muon example, Bob already `sees'
from the very beginning an electron and two neutrinos! This quite astonishing
result is a consequence of the fact that we `measure' if a state is decayed is
decayed or not in relation to the states with definite momentum $\left\vert
S,p\right\rangle $. In this context we refer also to Ref.
\cite{stefanovichnew} where this `decay via boost' was discussed. The physical
correctness of this choice is reinforced by the following facts:

(i) As noticed already in\ Ref. \cite{stefanovich,giunti,shirokovnew}, the
would-be survival probability of the state $\delta(v=0)^{-1/2}\left\vert
S,v\right\rangle $ is given in the Breit-Wigner limit by
\begin{equation}
\delta(v=0)^{-1}\left\langle S,v\left\vert e^{-iHt}\right\vert
S,v\right\rangle ^{2}=e^{-i\gamma\Gamma t}\text{ ,} \label{vsp}%
\end{equation}
i.e. with an \textit{increased} decay width $\gamma\Gamma$ instead of reduced
one. This is a quite counterintuitive result but, clearly, this is \emph{not}
what we measure. Indeed, there is nothing wrong in this expectation value, but
it simply does \textit{not} correspond to our measurements of
decayed/non-decayed moving particle. Just as mentioned before, the amplitude
$\left\langle \Psi\left\vert e^{-iHt}\right\vert \Psi\right\rangle $ is
\textit{not} what we are looking for.

(ii) Contrary to point (i), all physical results can be correctly described if
to be decayed or not is w.r.t. $\left\vert S,p\right\rangle $ (the standard
expression (\ref{gammatilde}) is re-obtained to an astonishingly good
approximation, see Fig. 1).

(iii) The momentum is a conserved quantity in\ QM and QFT and is naturally as
well as technically suited to serve as a preferred and physical basis
\cite{urbanowskicomment}.

In conclusion, if Alice prepares in her reference frame a normalized state
$\left[  \delta(v=0)\right]  ^{-1/2}\left\vert S,v=0\right\rangle =$ $\left[
\delta(p=0)\right]  ^{-1/2}\left\vert S,p=0\right\rangle $ (with
$\delta(v=0)/\delta(p=0)=M$), then for Bob the state is given by $\left[
\delta(v=0)\right]  ^{-1/2}\left\vert S,v\right\rangle .$ But -in Bob system-
this state has a negligibly small overlap with the basis of undecayed states
$\left\vert S,p\right\rangle ,$ i.e. it already consists of decayed particles.

\section{Spread in velocity (second possibility for a wave-packet)}

A state of definite velocity is also an idealization. Let us now consider a
wave-packet of the type:%
\begin{equation}
\left\vert \Phi\right\rangle =\int_{-1}^{+1}\mathrm{dv}C(v)\left\vert
S,v\right\rangle \text{ with }\int_{-1}^{+1}dv\left\vert C(v)\right\vert
^{2}=1\text{.} \label{vwp}%
\end{equation}
It is obtained from Eq. (\ref{theta}) by choosing $\beta(m,p)=m^{-1/2}%
\gamma^{-3/2}a_{S}(m)C(p/\sqrt{p^{2}+m^{2}})$, thus it is not in the
factorized form $B(p)a_{S}(m)$ as Eq. (\ref{pndpwp}); as a consequence,
$P_{nd}(0)<1.$\textbf{ }Explicitly,\textbf{ }the non-decay probability
amplitudes associated to $\left\vert \Phi\right\rangle $ are:
\begin{equation}
\left\langle S,p\left\vert e^{-iHt}\right\vert \Phi\right\rangle =\int
_{-1}^{+1}\mathrm{dv}C(v)d_{S}\left(  \frac{p}{\gamma v}\right)  \sqrt
{\frac{p}{v}}\frac{1}{v}e^{-i\gamma mt}%
\end{equation}
and the non-decay probability by:%
\begin{equation}
P_{nd}(t)=\int_{-\infty}^{+\infty}\mathrm{dp}\left\vert \left\langle
S,p\left\vert e^{-iHt}\right\vert \Phi\right\rangle \right\vert ^{2}.
\label{pndvwp}%
\end{equation}
In addition to $P_{nd}(0)<1$, one has also $P_{nd}(0)>0.$ The state
$\left\vert \Phi\right\rangle $ in Eq. (\ref{vwp}) is not a pure unstable
state, but there is a nonzero overlap with the basis of undecayed states
$\left\vert S,p\right\rangle $. Let us consider for illustration a Gaussian
form $C(v)=Ne^{-(v-v_{0})^{2}/(4\sigma_{v}^{2})}$ (where $N$ is such that the
normalization is fulfilled. To a very good approximation, for a not too wide
wave packet, one has $N=\left[  2\pi\sigma_{v}^{2}\right]  ^{-1/4}$).%

\begin{figure}
[ptb]
\begin{center}
\includegraphics[
height=2.3307in,
width=6.5942in
]%
{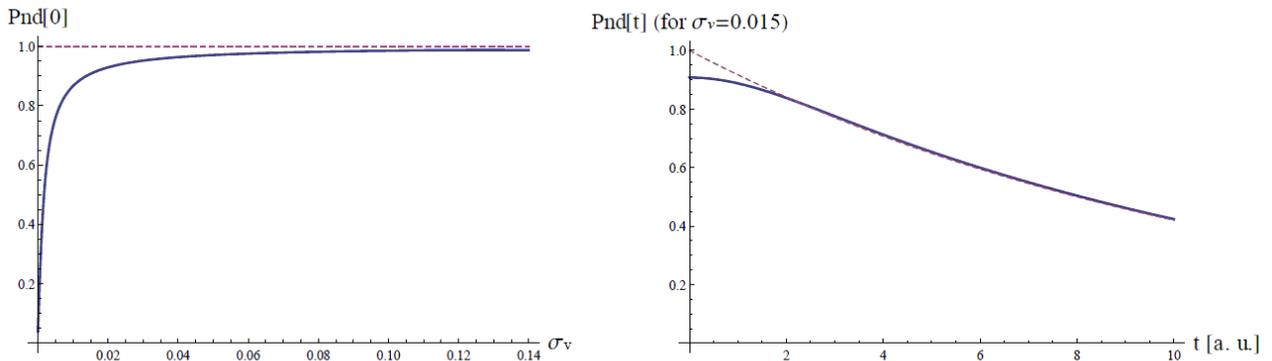}%
\caption{Left: $P_{nd}(0)$ as function of the velocity wave-packet spreading
parameter $\sigma_{v}.$ Right: $P_{nd}(t)$ for the illustrative case
$\sigma_{v}=0.015;$ after a transient non-exponential part, the standard
formula (\ref{gammatilde}) (dashed line) is a good approximation.}%
\end{center}
\end{figure}

In Fig. 2, left panel, we plot $P_{nd}(0)$ as function of the spreading
$\sigma_{v}$ for the particular choice $v_{0}=1/2$, $M=10$ [a. u.] and
$\Gamma/M=1/100.$ The smaller is the spreading, the smaller is also the
initial probability $P_{nd}(0)$. For $\sigma_{v}\rightarrow0$ we reobtain a
state with definite velocity and the overlap vanishes, as in\ Eq.
(\ref{noov}). However, for increasing $\sigma_{v}$ one gets very quickly a
state with $P_{nd}(0)\simeq1,$ which could be rewritten --to a very good
approximation-- in the form given by Eq. (\ref{pwp}). Thus, a state with
definite velocity is an idealization which must be regarded with great care.
In Fig. 2, right panel, we show also the behavior of $P_{nd}(t)$ (solid line)
for the representative case $\sigma_{v}=0.015,$ in which $P_{nd}(0)=0.9.$ A
part from a Zeno-like non exponential part at the beginning, it quickly
reaches the expected behavior (dashed line) $e^{-\Gamma_{p}t}\simeq
e^{-\tilde{\Gamma}_{p}t}=e^{-\Gamma t/\gamma_{0}}$ with $\gamma_{0}%
=(1-v_{0}^{2})^{-1/2}$, i.e. the standard formula is again a very good
approximation, as expected. In the end of this section, we emphasize that the
results depicted in Fig. 2 belong to the most important ones of the present manuscript.

\section{Discussions and conclusions}

\emph{ }The study of the decay law of fast moving unstable particles
represents an interesting link between special relativity and QM and QFT. This
is indeed an important subject which has received attention from valuable
pioneering works
\cite{exner,khalfin2,shirokovold,shirokov,shirokovnew,stefanovich,stefanovichnew,stefanovichbook,urbanowski}
but which also needs further clarifications and investigations in order to be
accepted by a larger part of the scientific community. Moreover, in view of
the very recent revival of the subject (e.g. Refs.
\cite{urbanowski,giunti,urbanowskinew}), we think that new interesting results
can be obtained in the near future in this interesting topic.

In the present work we have thus reviewed -especially in Sec. 2 and 4- some
existing results, but we have also presented new (in some cases analytic)
formulas on the subject. Moreover, in Sec. 3 and especially 5, we have
presented our main novel results. We briefly reports the main points in the
following paragraphs.

An unstable state with definite momentum $p$ has a decay width given by Eq.
(\ref{gammap}): it is slightly \emph{larger} than the usual expression of Eq.
(\ref{gammatilde}). Thus, the lifetime is actually slightly smaller than what
the time dilatation expression tells us. The standard \textquotedblleft
Einstein\textquotedblright\ formula remains however a very good approximation
which can be used in all practical applications, see Fig. 1 for the numerical
quantification of the deviation. It is an open question if the small deviation
will be measured at some point in the future.

More realistically, an unstable state is a superposition of the form of Eq.
(\ref{pwp}). The non-decay probability is not given by the probability that
the state is still the initial one, but one has to project onto the basis of
unstable states with definite momentum. The outcome is given in\ Eq.
(\ref{pndpwp}) and can usually also be very well approximated by the Einstein
formula. We also generalize the expressions to the completely arbitrary
initial state of Eq. (\ref{theta}), whose associated non-decay probability is
given by Eq. (\ref{pndtheta}). The various cases described in this work can be
obtained by taking proper limits of these general expressions.

A boost is a very subtle operation when applied to an unstable state of zero
momentum/velocity. Namely, the boosted state is an eigenstate of velocity but
has a zero overlap with the basis of undecayed states. Moreover, the
probability that such a state remains unchanged over time shows a quite
curious time-contraction feature, see Eq. (\ref{vsp}), but this is
\textit{not} the non-decay probability. When allowing for a realistic spread
in velocity, there is indeed a (usually large) overlap with the set of
undecayed states, and the non-decay probability can be again be well described
by the usual expression of Eq. (\ref{gammatilde}), as expected, see Fig. 2.

In conclusion, as usual in QM and QFT, one should specify which observer (if
Alice or Bob) is making the measurement. The reference frame of the detector
is relevant.

In this work, we gave expressions for the non-decay probability in the most
general form, but we did not discuss numerical examples in which deviations
from the Breit-Wigner spectral function appear. We did so because we wanted
first to focus on the exponential limit and because deviations are typically
small. Moreover, the use of the exponential limit is also advantageous because
the decay law is valid independently if and how many measurements are
performed \cite{koshinorev,gppra}. However, it must be stressed that the study
of non Breit-Wigner distributions is indeed very interesting, e.g. Ref.
\cite{urbanowski}. In this context, the study of numerical cases in which wave
packets of the most general type and the implementation of realistic
distributions is promising and constitutes an important outlook of the present
work. Moreover, the investigations of decays of moving particles at late times
might be of interest in cosmological particle physics because unstable
particles can be very fast and may travel long distances for a long time.

\bigskip

\textbf{Acknowledgments: }I thank G. Pagliara, W. Florkowski, W. Broniowski,
and J. R\'{o}g for useful discussions.


\begin{thebibliography}{99}                                                                                               %


\bibitem {schroedinger}Schr\"{o}dinger, Erwin (November 1935).
Naturwissenschaften 23 (49): 807--812 (1935).

\bibitem {ww}
V.~Weisskopf and E.~P.~Wigner,
Z.\ Phys.\ \textbf{63} (1930) 54.
V.~Weisskopf and E.~Wigner,
Z.\ Phys.\ \textbf{65} (1930) 18.
G. Breit, Handbuch der Physik 41, 1 (1959).

\bibitem {scully}M. O. Scully and M. S. Zubairy (1997), \textit{Quantum
optics}, Cambridge UK: Cambridge University Press.

\bibitem {khalfin}L. A. Khalfin, 1957 Zh. Eksp. Teor. Fiz. \textbf{33} 1371.
(Engl. trans. Sov. Phys. JETP \textbf{6} 1053).

\bibitem {misra}B.~Misra and E.~C.~G.~Sudarshan,
J.\ Math.\ Phys.\ \textbf{18} (1977) 756


\bibitem {dega}A.~Degasperis, L.~Fonda and G.~C.~Ghirardi,
Nuovo Cim.\ A \textbf{21} (1973) 471.


\bibitem {ghirardi}L.~Fonda, G.~C.~Ghirardi and A.~Rimini,
Rept.\ Prog.\ Phys.\ \textbf{41} (1978) 587.


\bibitem {koshinorev}K.~Koshino and A.~Shimizu,
Phys.\ Rept.\ \textbf{412} (2005) 191.


\bibitem {facchiprl}P.~Facchi, H.~Nakazato, S.~Pascazio
Phys.\ Rev.\ Lett.\ \textbf{86} (2001) 2699-2703.

\bibitem {zenoqft}F.~Giacosa, G.~Pagliara,
Mod.\ Phys.\ Lett.\ \textbf{A26 } (2011) 2247-2259. [arXiv:1005.4817
[hep-ph]];
F.~Giacosa and G.~Pagliara,
Phys.\ Rev.\ D \textbf{88} (2013) 025010 [arXiv:1210.4192 [hep-ph]].


\bibitem {duecan}F.~Giacosa,
Found.\ Phys.\ \textbf{42} (2012) 1262 [arXiv:1110.5923 [nucl-th]].




\bibitem {raizen}S.~R.~Wilkinson, C.~F.~Bharucha, M.~C.~Fischer,
K.~W.~Madison, P.~R. Morrow, Q.~Niu, B.~Sundaram, M.~G.~Raizen, Nature
\textbf{387}, 575 (1997);
M.~C.~Fischer, B.~Guti{\'{e}}rrez-Medina and M.~G.~Raizen, Phys. Rev. Lett.
\textbf{87}, 040402 (2001).


\bibitem {rothe}C. Rothe, S. I. Hintschich, A. P. Monkman, Phys. Rev. Lett.
\textbf{96} (2006)163601.

\bibitem {bailey}
J.~Bailey \textit{et al.},
Nature \textbf{268} (1977) 301.
J.~Bailey \textit{et al.} [CERN-Mainz-Daresbury Collaboration],
Nucl.\ Phys.\ B \textbf{150} (1979) 1.


\bibitem {lee}T.~D.~Lee,
Phys.\ Rev.\ \textbf{95 } (1954) 1329-1334.
C.~B.~Chiu, E.~C.~G.~Sudarshan and G.~Bhamathi,
Phys.\ Rev.\ D \textbf{46} (1992) 3508.


\bibitem {ford}P. R. Berman and G. W. Ford, Phys. Rev. A \textbf{82}, Issue 2,
023818; A. G Kofman, G. Kurizki, and B. Sherman,
Journal of Modern Optics, vol. 41, Issue 2, p.353-384 (1994); P. Facchi and S.
Pascazio, Phys. Rev. A \textbf{62}, 023804 (2000) [arXiv:quant-ph/9909043 ];
P.~Facchi and S.~Pascazio,
Phys.\ Lett.\ A \textbf{241} (1998) 139 [arXiv:quant-ph/9905017].
F.~Giacosa,
Phys.\ Rev.\ A 88, \textbf{052131} (2013) [arXiv:1305.4467 [quant-ph]].

\bibitem {exner}
P.~Exner,
Phys.\ Rev.\ D \textbf{28} (1983) 2621.


\bibitem {khalfin2}L. A. Khalfin, Quantum Theory of unstable particles and
relativity, PDMI Pewprint 6/1997.

\bibitem {shirokovold}M. I. Shirkokov,
JINR E2 10614 (1977).

\bibitem {shirokov}
M.~I.~Shirokov, Int.\ J.\ Theor.\ Phys.\ \textbf{43} (2004) 1541.


\bibitem {shirokovnew}M. I. Shirokov, arXiv:quant-ph/0508087

\bibitem {stefanovich}E.V. Stefanovich,
Internation l Journal of Theoretical Physics, Volume 35, Issue 12,
pp.2539-2554 (1996).

\bibitem {stefanovichnew}E. V. Stefanovich,
arXiv:physics/0603043

\bibitem {stefanovichbook}
E.~V.~Stefanovich,
\textquotedblleft Relativistic quantum dynamics: A Non-traditional perspective
on space, time, particles, fields, and action-at-a-distance,\textquotedblright%
\ physics/0504062 [physics.gen-ph].


\bibitem {urbanowski}
K.~Urbanowski,
Phys.\ Lett.\ B \textbf{737} (2014) 346 [arXiv:1408.6564 [hep-ph]].


\bibitem {giunti}
S.~A.~Alavi and C.~Giunti,
Europhys.\ Lett.\ \textbf{109} (2015) 6, 60001 [arXiv:1412.3346 [quant-ph]].


\bibitem {pdg}K. A. Olive et al. (Particle Data Group), Chin. Phys.
\textbf{C38}, 090001 (2014).

\bibitem {peskin}
M.~E.~Peskin and D.~V.~Schroeder, \textquotedblleft An Introduction to quantum
field theory,\textquotedblright\ Reading, USA: Addison-Wesley (1995) 842 p.


\bibitem {urbanowskicomment}
K.~Urbanowski,
arXiv:1504.00794 [quant-ph].

\bibitem {urbanowskinew}
K.~Urbanowski,
Adv.\ High Energy Phys.\ \textbf{2015} (2015) 461987.


\bibitem {gppra}
F.~Giacosa and G.~Pagliara,
Phys.\ Rev.\ A \textbf{90} (2014) 5, 052107 [arXiv:1405.6882 [quant-ph]].
\end{thebibliography}
\end{document}